\documentclass[a4paper,fleqn,usenatbib,useAMS]{mnras}

\usepackage{graphicx,multicol} 
\usepackage{times}
\def\etal{\it et~al.}

\title[Drifting and Periodic Nulling in PSR B2000+40]{Subpulse Drifting and Periodic Nulling in single pulse emission of PSR B2000+40}
\author[Basu, Lewandowski \& Kijak]{Rahul Basu$^1$, Wojciech Lewandowski$^2$, Jaros\l{}aw Kijak$^2$ \\
$^{1}$ Inter-University Centre for Astronomy and Astrophysics, Pune, 411007, India; rahulbasu.astro@gmail.com \\
$^{2}$ Janusz Gil Institute of Astronomy, University of Zielona G\'ora, ul. Szafrana 2, 65-516 Zielona G\'ora, Poland \\
}
\begin{document}



\maketitle

\label{firstpage}

\begin{abstract}
We have carried out a detailed study of single pulse emission from the pulsar 
B2000+40 (J2002+4050), observed at 1.6 GHz frequencies using the Effelsberg 
radio telescope. The pulsar has three components which are not well separated, 
with the central component resembling core emission. We have investigated 
modulations in single pulse behaviour using the fluctuation spectral analysis 
which showed presence of two prominent periodicities, around 2.5$P$ and 40$P$, 
respectively. The shorter periodicity was associated with the phenomenon of 
subpulse drifting and was seen to be absent in central core component. Drifting
showed large phase variations in conal components. Additionally, the periodic 
modulations had significant evolution with time, varying between very sharp and
highly diffuse features. In addition to drifting the pulsar also had presence 
of nulling in the single pulse sequence. The longer periodic feature in the 
fluctuation spectra was associated with nulling behaviour. The pulsar joins a 
select group which shows the presence of phase modulated drifting as well as 
periodic nulling in the presence of core emission. This provides further 
evidence for the two phenomena to be distinct from each other with different 
physical origin.
\end{abstract}


\section{Introduction}
\noindent
The single pulse emission in normal period pulsars ($P >$ 0.1 seconds) are 
characterised by the presence of different phenomena like subpulse drifting, 
nulling and mode changing \citep{dra68,bac70,bac73}. The pulsed radio emission 
is usually composed of one or more components called subpulses. In certain 
cases subpulses show systematic periodic variations within the pulse window and
the phenomenon is known as subpulse drifting. Nulling and mode changing on the 
other hand is seen as large scale variations in the radio emission 
\citep{wan07}. During mode changing the pulsar emission switches between more 
than one stable state with different profile shapes. While nulling corresponds 
to the condition where the stable emission switches off and goes below 
detection limit for varying durations lasting from few periods, to hours and 
even months at a time in the case of intermittent behaviour \citep{kra06}. An 
association between subpulse drifting and nulling behaviour was proposed with 
the discovery of periodic nulling \citep{her07,her09}. In certain pulsars it 
was observed that nulls, usually of short durations, were periodic in nature 
and co-existed with subpulse drifting whose periodicities were much shorter 
than nulling periodicity. This behaviour was explained using the carousel model
where subpulse motion was associated with a rotating sub-beam system 
\citep{gil00,des01}. Periodic nulls were interpreted as empty line of sight 
(LOS) traverse between the sub-beam system and were termed as pseudo-nulls, 
distinguishing them from the general nulling phenomenon.

A systematic study of periodic behaviour in the single pulse sequences of 
pulsar population have been carried out in the recent past by 
\citet{bas16,bas17,bas19a,bas20a}. A detailed classification of drifting 
behaviour revealed a close association with the profile type. The average radio
emission beam is expected to consist of a central core component surrounded by 
concentric rings of conal emission. The average profile shape is believed to be
closely associated with the LOS geometry across the emission beam 
\citep{ran93}. In case of tangential LOS traverses across the edge of the beam,
the conal single (S$_d$) profile is observed. As the LOS cuts more interior 
regions of the beam, profile shapes become more complicated ranging from conal 
double (D), conal triple ($_c$T) and conal quadruple ($_c$Q). For central LOS 
traverses the core dominated profiles, core single (S$_t$), triple (T) and 
multiple (M) are seen. Subpulse drifting is seen to vary greatly across the 
different profile classes. The systematic coherent drifting with prominent 
drift bands and large phase variations are usually associated with peripheral 
LOS traverse of the emission beam in S$_d$ and D profile classes. Drift 
behaviour becomes complicated for more interior LOS traverses, associated with 
$_c$T and $_c$Q profiles, where 180\degr~phase jumps and phase reversals are 
seen across adjacent profile components. In the case of central LOS traverse, 
firstly, no drifting is seen in central core component, and secondly, in conal 
components prominent drifting behaviour is observed which usually have low 
phase variations \citep{bas19a}. 

These studies also reveal distinctions between subpulse drifting and other 
periodic phenomena like periodic amplitude modulation and periodic nulling seen
in the single pulse sequence. In contrast to subpulse drifting periodic 
amplitude modulations and periodic nulling are not restricted to the subpulse 
behaviour, but are seen across the entire profile. Subpulse drifting is seen in
pulsars with spin-down energy loss ($\dot{E}$) less than 2$\times$10$^{32}$
erg~s$^{-1}$, and their periodicities are weakly anti-correlated with $\dot{E}$
\citep{bas16,bas19a}, i.e. pulsars with lower $\dot{E}$ tend to have higher 
drifting periodicities and vice versa. On the other hand periodic nulling and 
periodic amplitude modulation are seen over a wide $\dot{E}$ range, much higher
than the drifting boundary, and do not show any dependence on $\dot{E}$ 
\citep{bas17,bas20a}. This has prompted the assertion that they are a newly 
emergent phenomenon in pulsars distinct from subpulse drifting \citep{bas17,
bas20a}. Additionally, a clear distinction is seen in relation to the behaviour
of core emission. As noted earlier subpulse drifting is not seen in the core, 
while the periodic modulations are seen across the entire profile including the
core. There are only three pulsars with prominent core emission which show the 
presence of both subpulse drifting as well as periodic modulations in their 
pulse sequence, PSR J1239+2453 \citep{smi13,maa14,bas20a}, PSR J1740+1311 
\citep{for10} and PSR J2006$-$0807 \citep{bas19b}. All the three pulsars have M
profile type, with PSR J1239+2453 and J2006$-$0807 showing periodic nulling, 
while PSR J1740+1311 exhibit periodic amplitude modulation. 

The above discussion signifies the importance of studying periodic behaviour in
pulse sequences particularly in presence of a central core component. The 
primary focus of this work is to carry out detailed analysis of single pulse 
behaviour of the pulsar B2000+40 (J2002+4050) which has three components, with 
the central component resembling core emission. The presence of subpulse 
drifting at 21~cm observing wavelength was reported for this source by 
\citet{wel06}. However, no drifting behaviour could be seen at 92~cm 
observations of \citet{wel07} as well as the 333 MHz studies of \citet{bas19a}.
The low radio frequency observations had lower detection sensitivities of 
single pulses and were likely affected by scattering effects in the intervening
medium. There were indications of low frequency features corresponding to 
periodic intensity modulations in the 333 MHz observations, but they could not 
be confirmed. This prompted us to carry out a detailed study of this source at 
a higher frequency. We have carried out a thorough analysis of the radio 
emission from this pulsar. In section \ref{sec:obs} we present the 
observational setup and analysis scheme used to generate the single pulse 
sequence. Section \ref{sec:avgprof} details the behaviour of the average 
profile while sections \ref{sec:permod} and \ref{sec:null} presents the single 
pulse analysis leading to the characterisation of subpulse drifting and 
nulling, respectively. A discussion of the implications of the results of this 
study is presented in section \ref{sec:disc} followed by a short summary and 
conclusions in section \ref{sec:sum}.

\section{Observation and Analysis}\label{sec:obs}

The data presented here is a part of an observing project that involved single 
pulse observations of eight strong northern hemisphere pulsars. The 
observations were performed with the 100-meter Effelsberg radio telescope 
between March 9 and March 11, 2018. Data was acquired using the L-band receiver
with the central frequency of 1635~MHz with a bandwidth of 250~MHz. Each pulsar
was observed for a minimum of ten thousand individual pulses. The Effelsberg 
pulsar recording machine PSRIX \citep{lazarus16} was used to record coherently 
dedispersed data and then to translate it into 128 frequency channels and 
between 1024 and 4096 phase bins depending on a pulsar period (2048 bins in the
case of PSR J2002+4050), in a single-polarization mode (i.e. total intensity 
only). Data were saved in the {\sc Archive} format for each pulse separately 
and then merged to create a single file.

The next steps of the analysis, including the interference (RFI) cleaning 
process was performed with the help of the {\sc Psrchive} package 
\citep{Hotan2004}. The narrow band interference was removed using a median 
filter as implemented by the ``zap'' routine in {\sc Psrchive} 
\citep{vanStraten2012}.  Due to heavy interference we were forced to remove 
more than a half of the spectral channels, reducing the effective bandwidth to 
about 100~MHz. Initial 30 percent and around 10 percent of the bandwidth at the
end was removed due to RFI. Additional, 10-20 percent of spectral channels 
distributed within the remaining band were also affected by narrow band RFI. 
The effective central frequency was around 1660 MHz. After the interference 
removal the data was dedispersed to create the total intensity time series and 
translated into the ASCII format for further processing.

\section{Average Profile Behaviour}\label{sec:avgprof}

\begin{figure}
\begin{center}
\includegraphics[scale=0.68,angle=0.]{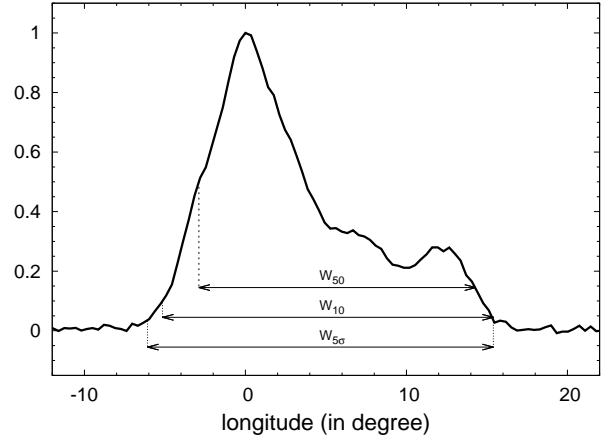}
\end{center}
\caption{The figure shows average profile of PSR J2002+4050 at 1.6 GHz. The 
profile shows presence of three components, with the leading conal component 
being the dominant one and not clearly separated from the central core 
emission. The boundaries corresponding to the profile widths W$_{5\sigma}$, 
W$_{10}$ and W$_{50}$ are also shown n the figure.}
\label{fig_prof}
\end{figure}

\begin{table}
\resizebox{\hsize}{!}{
\begin{minipage}{80mm}
\caption{Measuring Profile widths of PSR J2002+4050}
\centering
\begin{tabular}{cccc}
\hline
 W$_{5\sigma}$ & W$_{10}$ & W$_{50}$ & W$^{cone}_{SEP}$ \\
 (\degr) & (\degr) & (\degr) & (\degr) \\
\hline
 21.5$\pm$0.4 & 20.6$\pm$0.4 & 17.2$\pm$0.4 & 12.1$\pm$0.4 \\
\hline
\end{tabular}
\label{tabwid}
\end{minipage}
}
\end{table}

\noindent
We have estimated the average profile of the pulsar at 1.66 GHz observing 
frequency which is shown in figure \ref{fig_prof}. The profile shows presence 
of three components, with the leading component being the most prominent having
peak intensity more than three times the other two components. The different 
components in the profile are not clearly separated at the observing frequency 
with the central component being merged with the leading one. In Table 
\ref{tabwid} we have estimated the profile widths as well as separation between
the peaks of the two conal components (W$^{cone}_{SEP}$). The peak location of 
the leading component was seen in the average profile. However, the peak 
location of the trailing component was not clearly identified and the centroid 
was used. Three separate estimates were carried out for the profile width as 
shown in figure \ref{fig_prof}. W$_{5\sigma}$ was the separation between the 
two longitudes at the leading and trailing edge of the profile that were five 
times the baseline noise rms level. Additionally, we also estimated W$_{10}$ 
and W$_{50}$, which corresponded to separation between 10 percent peak 
intensity level of the leading and trailing components and the 50 percent peak 
intensity levels of the two components. We have not come across any clear 
classification of this pulsar profile in the literature. The observational 
setup did not allow us to estimate polarization behaviour. However, the 
polarization in the average profile has been reported in earlier works 
\citep{gou98}. The polarization position angle suggests large swing across the 
pulsar profile consistent with a central line of sight traverse \citep{rad69}. 
There are also indications of orthogonal polarization modes which has important
implications for understanding the emission mechanism \citep{mit09,mel14}. The 
linear polarization shows depolarization effect towards the profile edges while
the circular polarization show sign changing behaviour near the central 
component. The above features are consistent with core-cone T profile type. The
drifting behaviour reported in section \ref{sec:permod} provides additional 
support for this profile classification.

\section{Periodic modulations of single pulses}\label{sec:permod}

\begin{figure*}
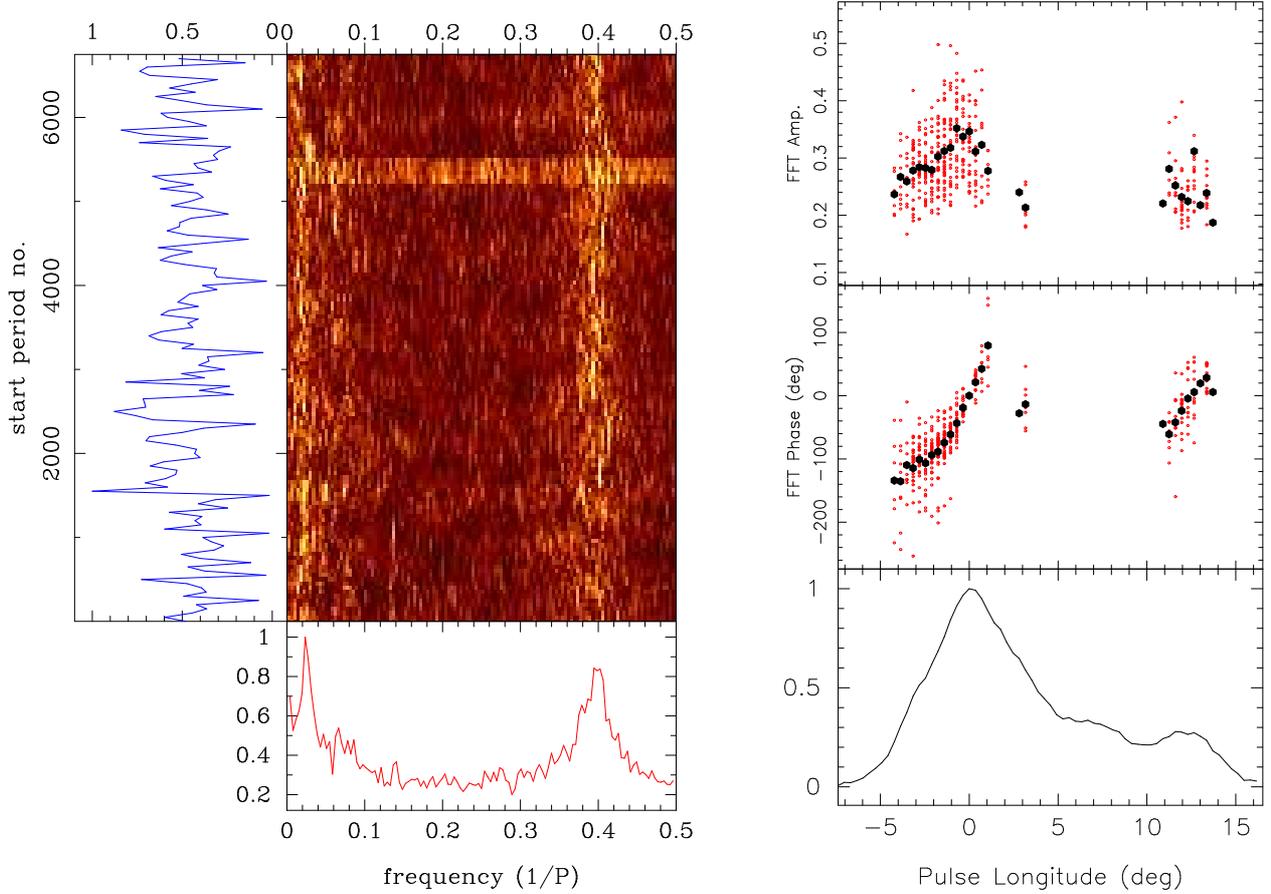

\begin{tabular}{@{}cr@{}}
{\mbox{\includegraphics[scale=0.47,angle=0.]{B2000+40_LRFSavg.ps}}} &
\hspace{20px}
{\mbox{\includegraphics[scale=0.47,angle=0.]{B2000+40_peakphs.ps}}} \\
\hspace{30px}
\end{tabular}
\caption{The figure shows fluctuation spectral analysis on single pulse 
sequence of PSR J2002+4050. The left panel shows time variation of the 
Longitude resolved fluctuation spectra (LRFS). The bottom window represents 
average LRFS across all times and shows presence of two clear peaks. The low 
frequency peak corresponds to periodic nulling while the high frequency peak
arises due to subpulse drifting. The right panel shows variation of the 
drifting behaviour across the pulse profile. The top window shows the variation
of the peak drifting amplitude while the middle window represents the average 
phase behaviour corresponding to drifting peak. Subpulse drifting is absent in 
the central core component. The conal components on the other hand show large 
phase variations.}
\label{fig_drift}
\end{figure*}

\noindent
We have used the fluctuation spectral analysis to estimate periodic modulations
in the single pulse sequence. The details of the analysis is reported in 
\citet{bas18a}, which is briefly summarized as follows. The longitude resolved 
fluctuation spectra \citep[LRFS,][]{bac73} was estimated for 256 consecutive 
periods. The LRFS was estimated at regular intervals, from the beginning of 
observations and subsequently shifting the starting pulse by fifty periods. The
average LRFS amplitude, across all longitudes along the pulse window, was 
represented as a function of the starting period to determine the time 
evolution of periodic modulations. The time evolution of LRFS is shown in 
figure \ref{fig_drift}, left panel, which shows the presence of two distinct 
peaks in the fluctuation spectra. In addition to the time evolution of the 
amplitude, averaged for all longitudes, we have also estimated the variation of 
periodic behaviour across the pulse window. This was determined for the high 
frequency feature associated with subpulse drifting. The distribution of peak 
fluctuation amplitude at each longitude is shown in figure \ref{fig_drift}, 
top window of right panel (red points), along with the average behaviour (black
points). Additionally, we have also estimated the distribution of relative 
phase variations corresponding to the drifting peak, which is shown in the 
middle window of the plot. The relative phase variations were estimated for 
each LRFS, with the phases corresponding to the profile peak fixed at zero. The
peak amplitudes and phases were represented for significant detections, where 
the drifting peak exceeded three times the rms level of the baseline. The 
single pulse sequence during our observations were affected by RFI which were 
likely to introduce artificial peaks in the fluctuation spectra. To mitigate 
such behaviour we identified a window of same width as the profile in the 
off-pulse region and estimated the LRFS corresponding to this window. The RFI 
being terrestrial in origin was present throughout the profile and not just in 
the pulse window. Subsequently, the LRFS in the off-pulse window was subtracted
from the LRFS of pulse window which suppressed the peaks arising due to RFI. In
none of these cases the RFI peaks were coincident with the modulation 
periodicities of the pulsar emission.

\begin{figure*}
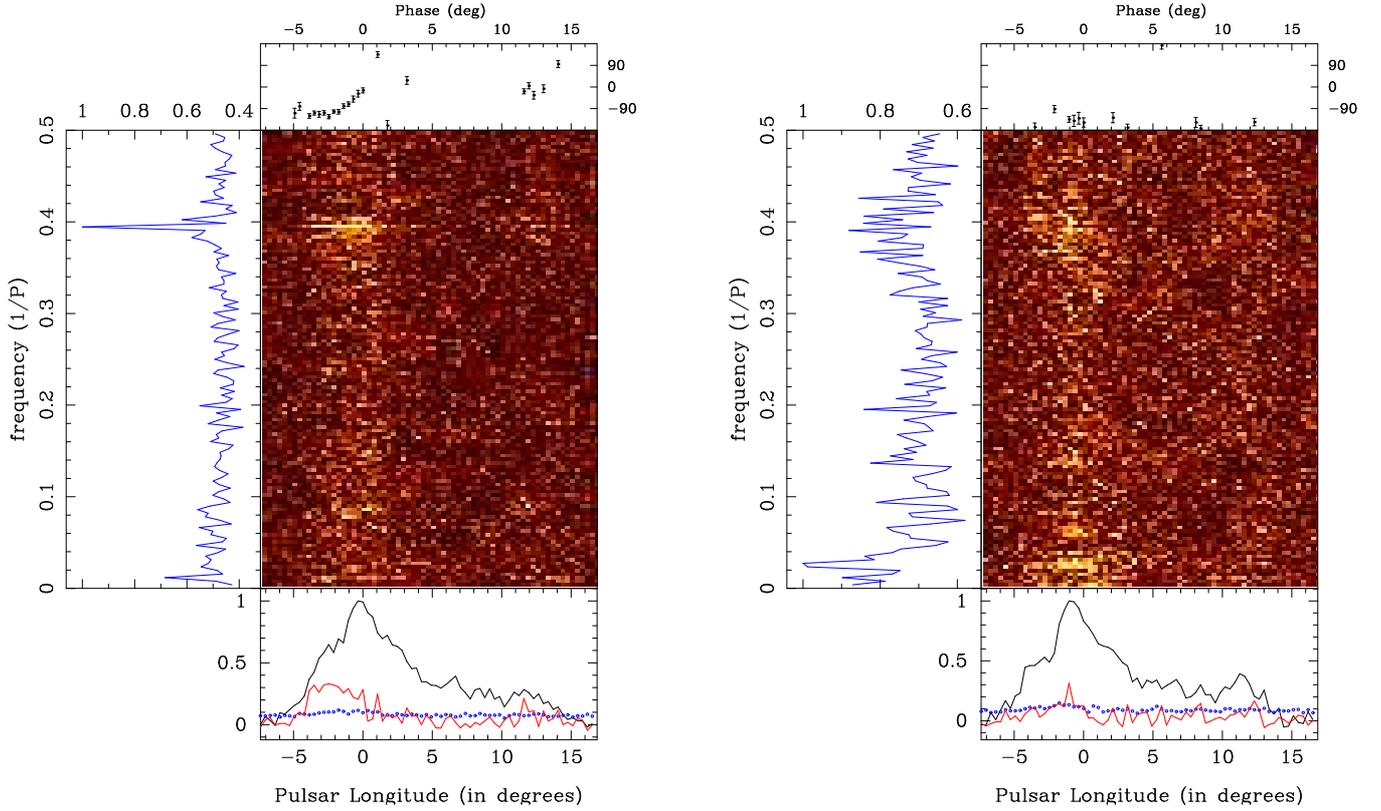

\begin{tabular}{@{}cr@{}}
{\mbox{\includegraphics[scale=0.43,angle=0.]{B2000+40_lrfs_sharp.ps}}} &
\hspace{20px}
{\mbox{\includegraphics[scale=0.43,angle=0.]{B2000+40_lrfs_wide.ps}}} \\
\hspace{30px}
\end{tabular}
\caption{The figure shows variation of LRFS for the pulsar J2002+4050 during 
two intervals within the pulse sequence. The left panel corresponds to interval
between pulse 4000 and 4256 and shows presence of a sharp feature corresponding
to subpulse drifting. The right panel corresponds to interval between pulse 
1000 and 1256 and shows presence of diffuse drifting. The low frequency 
modulations on the other hand are more prominent in this interval.}
\label{fig_driftvar}
\end{figure*}

As shown in figure \ref{fig_drift}, left panel, there are two distinct periodic
features in the single pulse sequence. The average periodic behaviour does not 
appear to be very sharp in both cases and show significant variations with 
time. This appears to be a general behaviour of subpulse drifting as well as 
other periodic modulations, as noted in a number of other pulsars \citep{bas17,
bas18a}. These variations are highlighted in the LRFS estimated for the two 
specific intervals shown in figure \ref{fig_driftvar}. The left panel 
corresponds to the sequence between pulse number 4000 and 4256 from the start 
of the observations. The LRFS shows presence of a very sharp peak corresponding
to subpulse drifting, but the low frequency feature is very weak in this 
sequence. The right panel shows the LRFS corresponding to the sequence between 
pulses 1000 and 1256. The drifting behaviour is seen as a diffuse structure in 
this duration while the low frequency behaviour becomes much more prominent. 
Table \ref{tabper} reports estimates of peak features corresponding to the two 
phenomena from time average fluctuation spectra. The Table shows peak frequency
($f_p$), full width at half maximum (FWHM) of each feature, strength of each 
peaks (S$_M$) given by the maximum height from baseline level divided by FWHM, 
and corresponding periodicities ($P_M$). The error in estimating peak frequency
is obtained using a Gaussian approximation, $\delta f_p$ = 
FWHM/2$\sqrt{2\ln{2}}$ \citep{bas16}. 

\begin{table}
\caption{Estimating periodic behaviour in PSR J2002+4050}
\centering
\begin{tabular}{c@{\hskip3pt}c@{\hskip3pt}c@{\hskip3pt}c@{\hskip3pt}c@{\hskip3pt}c@{\hskip3pt}c}
\hline
   & $f_p$ & FWHM & S$_M$ & $P_M$ & \multicolumn{2}{c}{$d\phi/d\psi$} \\
   &  &  &  &  & COMP-1 & COMP-3 \\
\hline
   & (cy/$P$) & (cy/$P$) & ($P$/cy) & ($P$) & (\degr/\degr) & (\degr/\degr) \\
\hline
 Drift. & 0.396$\pm$0.018 & 0.042 & 12.6 & 2.52$\pm$0.11 & 41.1$\pm$3.0 & 27.9$\pm$3.0 \\
   &  &  &  &  &  &  \\
 Per. Null & 0.025$\pm$0.007 & 0.017 & 41.3 & 39.6$\pm$11.6 & ... & ... \\
\hline
\end{tabular}
\label{tabper}
\end{table}

The low frequency feature around periodicity of 40 $P$ corresponds to periodic 
nulling and is explored in detail in section \ref{sec:null}. The high frequency
feature with $P_3$ around 2.5 $P$ arises due to subpulse drifting. As seen in 
figure \ref{fig_drift}, right panel, subpulse drifting is only seen in the 
conal components and is completely absent in the central core emission. In the 
first conal component drifting is seen primarily in the leading part and is
greatly diminished near the trailing edge. The leading component is not clearly 
separated from the core emission and it is possible that the trailing part of 
the component is affected by the lack of drifting in the core, thereby 
curtailing the drifting behaviour. Subpulse drifting in conal components are 
also characterised by large phase variations. This is not usual for central LOS
traverses of the emission beam which are mostly accompanied by low phase 
variations \citep{bas19a}. We have estimated rate of variations of drift phase 
($\phi$) with the pulse longitude ($\psi$), which is represented as 
$d\phi/d\psi$ in Table \ref{tabper}, by linear approximation of the phase 
variations in each conal component. The leading conal component shows larger 
drift phase variations compared to the trailing component. The drift rate can 
also be estimated using this measurement as $D_R$ = 360\degr/($P_3\times 
d\phi/d\psi$). The average drift rate is 3.5$\pm$0.4 \degr/$P$ for the leading 
conal component and 5.1$\pm$0.8 \degr/$P$ for the trailing component.

\section{Nulling}\label{sec:null}

\begin{figure}
\begin{center}
\includegraphics[scale=0.54,angle=0.]{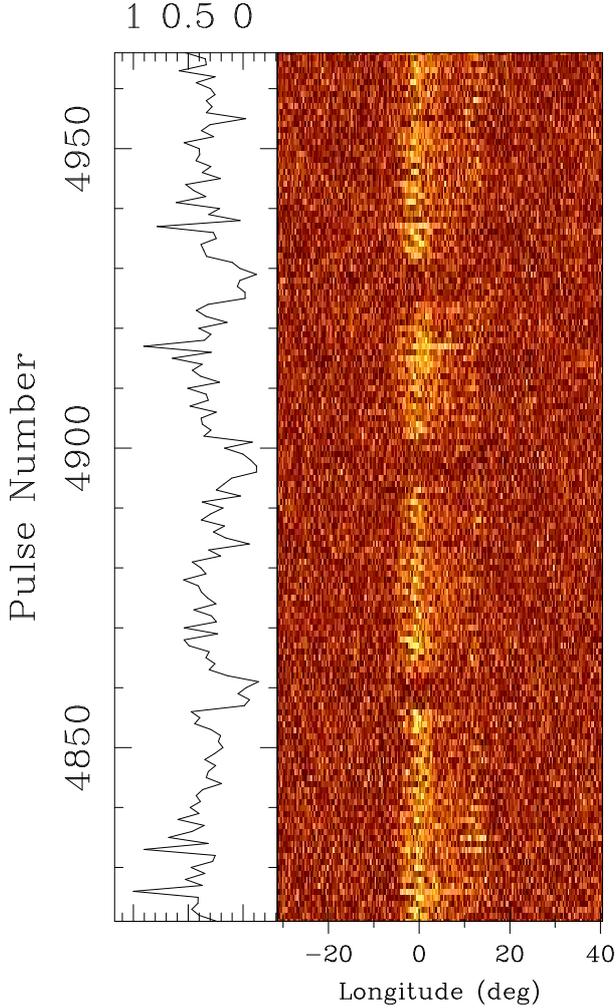}
\end{center}
\caption{The figure shows a section of single pulse sequence of pulsar 
J2002+4050, between pulse number 4820 and 4970 from the start of the observing
session. The pulsar shows presence of nulling at regular intervals. The figure 
shows presence of three prominent nulling regions around pulse 4860, 4900 and 
4930. The left window shows average energy corresponding to each single pulse, 
where the peak energy is normalized to unity. The energy drops to baseline 
noise levels during nulling intervals.}
\label{fig_nullsingl}
\end{figure}

\begin{figure}
\begin{center}
\includegraphics[scale=0.67,angle=0.]{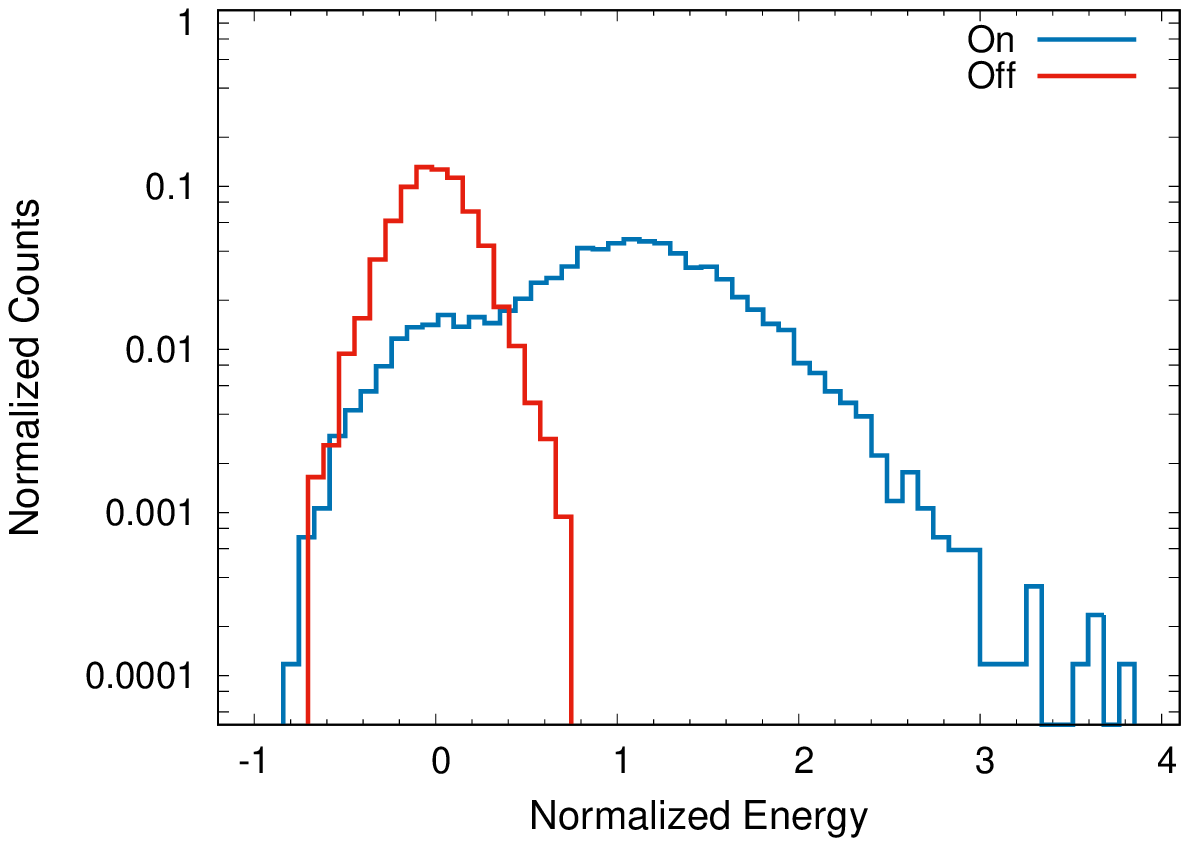}
\end{center}
\caption{The figure shows the pulse energy distribution corresponding to 
on-pulse (blue) and off-pulse (red) window. The on-pulse distribution shows a 
bimodal structure, where the lower part of distribution is coincident with 
off-pulse histogram. This indicates the presence of nulling in the pulsar 
J2002+4050.}
\label{fig_energydist}
\end{figure}

\noindent
We have detected the presence of nulling in the single pulse sequence of this 
pulsar. In figure \ref{fig_nullsingl} a short interval of the pulse sequence,
around 150 pulses, is shown which reflects the nulling behaviour. The figure 
shows that the pulsar goes to null state where the radio emission vanishes for 
short durations lasting between 5-10 $P$ at a time. In order to characterise 
the nulling behaviour we have determined the pulse energy distribution which is
shown in figure \ref{fig_energydist}. The average energy in the pulse window 
was estimated for each single pulse along with baseline average in a suitably 
selected off-pulse window. The histograms of the energy distributions were 
estimated in each case after normalizing the x-axis with average on-pulse 
energy and the y-axis with total number of pulses. The pulse energy 
distribution shows presence of a bimodal structure with lower part 
corresponding to null pulses. The nulling fraction was determined by fitting 
Gaussian functions to off-pulse and null distributions and estimating the ratio
between the two \citep{rit76}. The pulsar was in the nulling state for 
11.5$\pm$0.4 percent of time during the duration of our observations. One of 
the drawbacks of these estimates is the relatively low detection sensitivity of
the single pulse emission. This is also highlighted in the pulse energy 
distribution where null and burst distributions are not well separated. As a 
result it was impossible to identify the null pulses statistically and carry 
out a more detailed analysis of the nulling behaviour, like estimating the null
and burst length distribution, as well as searching for any low level emission 
during nulling.

\begin{figure*}
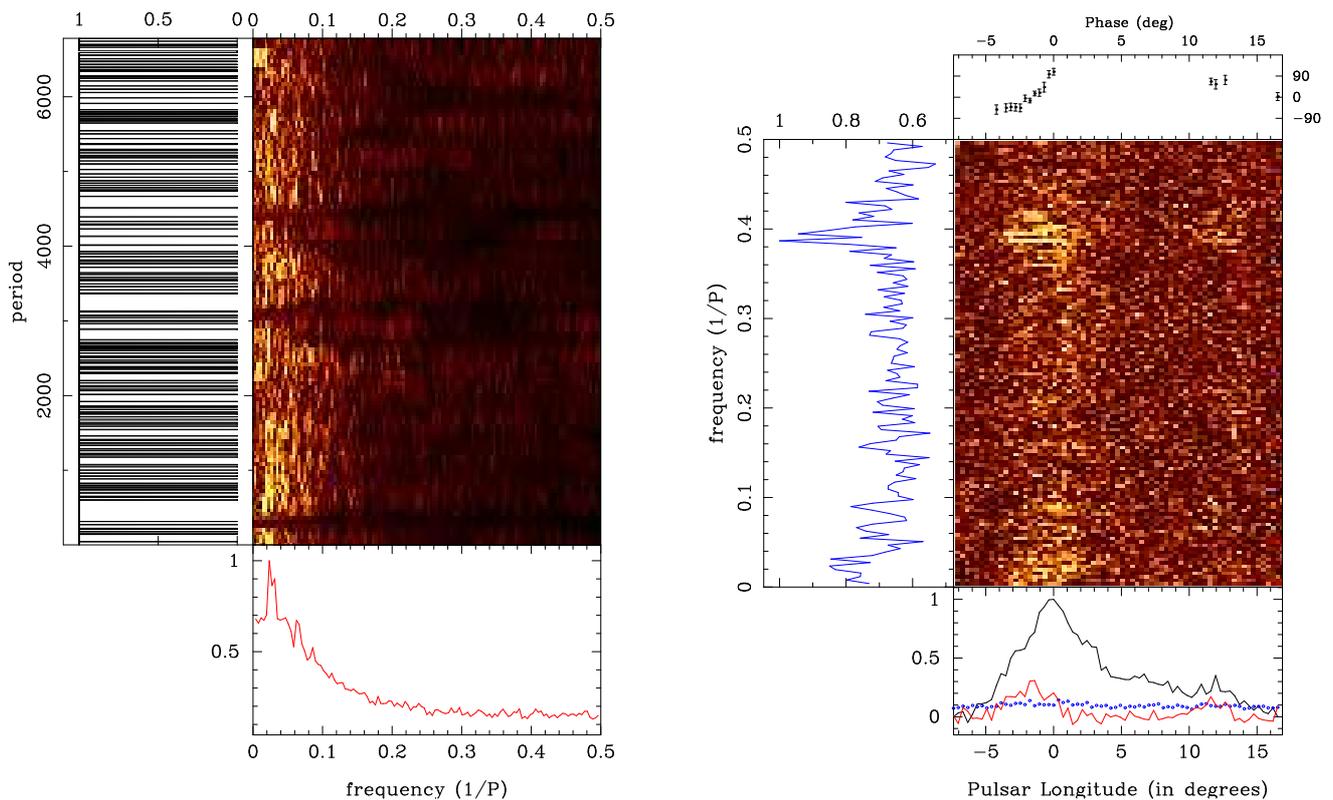

\begin{tabular}{@{}cr@{}}
{\mbox{\includegraphics[scale=0.42,angle=0.]{B2000+40_nullfft.ps}}} &
\hspace{20px}
{\mbox{\includegraphics[scale=0.42,angle=0.]{B2000_lrfs_nonull.ps}}} \\
\hspace{30px}
\end{tabular}
\caption{The left panel of the figure shows time evolution of FFT corresponding
to the null/burst time series. The nulls and burst pulses were identified as 0 
and 1, as shown in the left window, and a continuous FFT of the sequence was 
measured. The average FFT over the entire sequence is estimated in the bottom
window which shows the presence of a low frequency feature around 40 $P$. The 
peak frequency is coincident with the low frequency feature in the average 
fluctuation spectra. The right panel in the figure corresponds to the 
fluctuation spectra between pulse 3100 and 3400 where very little clear nulls 
are seen (see left panel). The absence of clear low frequency feature in the 
LRFS further suggests the longer periodicity modulation to be due to periodic 
nulling.}
\label{fig_nullfft}
\end{figure*}

One of the features of the fluctuation spectral analysis reported in the 
previous Section is the presence of low frequency feature around 0.025 cy/$P$. 
The presence of nulling gives rise to the possibility that the feature is 
indicative of periodic nulling behaviour in the pulsar. Although, low 
sensitivity of single pulse detection makes it impossible to identify the null 
and burst pulses statistically, figure \ref{fig_nullsingl} shows that it is 
still possible to identify null regions by visually inspecting the pulse 
sequence. We have employed this technique to largely identify the null and 
burst regions and carry out a periodicity analysis similar to \citet{bas17}. 
This involved setting up a binary sequence where the null pulses were 
identified as `0' and the burst pulses as `1'. Subsequently, Fourier transforms
of this series were carried out, similar to the fluctuation spectral analysis, 
using 256 points at a time, and continuously shifting the start period by 50 
pulses. The time evolution of the nulling FFT is shown in figure 
\ref{fig_nullfft}, left panel, where the average spectra shows the presence of 
a low frequency peak at $f_p$ = 0.027 cy/$P$. Firstly, the above analysis 
clearly demonstrates nulling to be periodic in nature, and secondly, the peak 
frequency in nulling FFT is coincident with the low frequency feature in 
fluctuation spectra indicating a common origin for both. As a result we can
conclusively ascertain that the longer periodic modulations detected in the 
single pulse sequence of this pulsar is indeed due to periodic nulling. To 
further demonstrate the validity of this claim we show the fluctuation spectra
in a region of the pulse sequence, between pulse number 3100 and 3400, where 
relatively fewer nulls were visible. The low frequency feature is not seen in 
LRFS as expected. The periodic behaviour is also indicated in the pulse 
sequence in figure \ref{fig_nullsingl}, where the interval between onset of 
nulls vary between 30 to 40 $P$ which matches with the measured periodicity in 
the fluctuation spectra. However, it should also be noted that the nulling FFT 
shown in figure \ref{fig_nullfft} is not an exact representation of the 
periodic behaviour since a number of null pulses, particularly single period 
nulls, were not likely to be identified due to the limitations of these 
observations, which may result in suppression of high frequency features in the
nulling FFT.

\section{Discussion}\label{sec:disc}
\noindent
The pulsar J2002+4050 joins a small, but increasing, group of sources which 
show the presence of both periodic nulling as well as subpulse drifting in 
their pulse sequences. A survey of such sources has been conducted recently by
\citet{bas20a}, where around 20 pulsars are reported to show the presence of 
both low frequency periodic modulation and subpulse drifting. Amongst them the
more relevant cases are the ones where there is a central core emission. There 
are only three such pulsars, PSR J1239+2453, J1740+1311 and J2006$-$0807, all 
of which have a five component M type profile. In case of the pulsars 
J1239+2453 \citep{bas20a} and J2006$-$0807 \citep{bas19b} the low frequency 
feature corresponds to periodic nulling while in PSR J1740+1311 \citep{for10} 
there are few nulls and the pulsar exhibit periodic amplitude modulation. PSR 
J2002+4050 is the first known example of a core-cone T profile where the conal 
components show the presence of subpulse drifting in addition to the presence 
of periodic nulls. In this context it is interesting to explore the periodic 
behaviour in PSR J2002+4050. 

As noted earlier subpulse drifting in PSR J2002+4050 shows large phase 
variations which are not common in pulsars with central LOS traverses, where 
generally flat phase variations are seen \citep{bas19a}. However, there are 
examples of large phase variations associated with certain conal emission in 
some cases. The pulsar J1239+2453 \citep{maa14,smi13,bas19a} shows large phase 
variations in the second conal component while the first and fifth components
show largely phase stationary behaviour. The drifting in the fourth component 
is weaker and the phase behaviour is not clearly seen. Similarly, in the case 
of PSR J2006$-$0807 the outer conal components show phase stationary behaviour 
with very little change, while the inner cones show large phase variations. 
However, the phase variations in the inner cones show the interesting 
phenomenon of bi-drifting where the drift directions, and consequently the 
phase slopes, are opposite in nature. The large phase variations seen 
associated with the drifting in PSR J2002+4050 seem to indicate that the conal 
components resemble the inner cones of the M type profiles. This is further 
highlighted by the fact that the components are not clearly separated. However,
unlike PSR J2006$-$0807, the phase variations in the two components are not in 
the opposite directions, despite, the slope in the leading component being much
steeper than the trailing component. 

The coherent radio emission originates due to non-linear instabilities in 
non-stationary, relativistic, outflowing plasma clouds moving along the open 
magnetic field lines \citep{ass98,mel00,mit09,lak18,rah20}. The non-stationary 
plasma clouds are generated due to sparking discharges in an inner acceleration
region (IAR) above the stellar surface, and requires presence of non-dipolar 
magnetic fields \citep{rud75,gil00,mit20}. The sparks undergo variable 
$\mathbf{E}\times\mathbf{B}$ drift due to presence of large electric and 
magnetic fields in IAR which is imprinted in the outflowing plasma and seen as
subpulse drifting in the radio emission. The phase variations seen associated 
with subpulse drifting in central LOS pulsars are particularly challenging to 
understand from the perspective of \citet{rud75} model of sparks rotating 
around the dipolar magnetic axis. Subsequent studies have introduced the 
concept of elliptical emission beams with axes tilted with respect to the 
fiducial plane in order to explain these phase variations \citep{wri17}. The 
physical origin of these tilted structures were explained by \citet{sza17,
sza20} by introducing the effect of non-dipolar fields in the polar cap which 
shifts it from the canonical dipolar location. It was suggested that the plasma
in the IAR rortate around a point of maximum potential at the polar cap. 
However, \citet{bas20b} showed that in the absence of any external electric 
field, the variable $\mathbf{E}\times\mathbf{B}$ drift in the IAR causes the 
sparks to lag behind the co-rotation motion of the pulsar, which is around the 
rotation axis. The non-dipolar polar cap is shifted from purely dipolar case, 
but in the emission region, around heights of few hundred kilometers from the 
surface, the magnetic field is purely dipolar in nature \citep{kij97,kij03,
mit02,krz09,mit17}. As the non-dipolar magnetic field line in the polar cap 
surface connects with the dipolar fields in the emission region, the field 
lines get twisted and the LOS traverses the sparks at different angles 
depending on the nature of the surface field. In case of pulsars like 
J2002+4050 with a central core component and large drifting in the cone, it was
found that the non-dipolar polar cap located on the sides of the neutron star 
relative to the dipolar case, i.e. roughly 90\degr~away, is a likely 
configuration for the drifting behaviour. However, more detailed modelling is 
required to explain the drifting phase behaviour in J2002+4050, particularly 
the different drift rates in the two conal components, etc. Other examples of 
similar behaviour needs to be uncovered in the pulsar population to find 
additional constraints on the physical models explaining the subpulse drifting 
in central LOS traverse systems.
 
A number of studies in the recent past \citep{bas16,mit17a,bas17,bas18b,bas19a,
bas19b,bas19c,bas20a,yan19,yan20} have highlighted the physical differences 
between subpulse drifting and low frequency modulations corresponding to 
periodic nulling and periodic amplitude modulation. The subpulse drifting is 
only associated with conal components while the entire pulsar emission is 
simultaneously affected by periodic modulations. The differences in the 
behaviour is clearly reflected in the single pulse emission of PSR J2002+4050 
where the subpulse drifting is absent in the central core component. The core 
component is merged with the leading conal component, whose trailing edge also 
seems to be affected by the core where drifting effect is significantly 
reduced. Nulling is not clearly distinguishable from low S/N single pulses, 
except during longer ($\sim$5 to 10 P) nulling intervals where the radio 
emission from all three components appears to vanish. It is possible that the 
pulsar goes to a low level emission state similar to PSR J1048$-$5832 
\citep{yan20}, rather than a complete null, which will require more sensitive 
future observations to uncover. The two periodic phenomena also differ in their
dependence on the spin-down energy loss ($\dot{E}$). The subpulse drifting is 
only seen in pulsars with $\dot{E} <$ 2$\times$10$^{32}$~erg and the drifting 
periodicity is weakly anti-correlated with $\dot{E}$, implying that for higher 
values of $\dot{E}$ the $P_3$ is generally lower. The best fit values to the 
$P_3 - \dot{E}$ correlation was estimated by \citet{bas16} as $P_3 = 
(\dot{E}/2.3\times10^{32})^{-0.6}$. The pulsar J2002+4050 has $\dot{E}$ of 
9.26$\times$10$^{31}$ erg~s$^{-1}$ \citep{hob04}, which is near the upper 
boundary but does not exceed it. The estimated $P_3$ of 2.5$P$ is relatively 
low and the corresponding aliased value $P_3^a$ is 1.66$P$. The expected $P_3$
from the above correlation is 1.72$P$ which suggests the aliased periodicity is
a better fit to the correlation. However, there is large scatter in the $P_3 - 
\dot{E}$ distribution which makes such associations tentative. On the other 
hand the low frequency periodic modulations do not show any clear dependence on
$\dot{E}$ and have periodicities which appear in a relatively wide band with 
$P_M$ between 10-200$P$ \citep{bas20a}. The periodic nulling in PSR J2002+4050 
has $P_M$ of around 40$P$ which is also consistent with expected behaviour. The
periodic modulations observed in the single pulse sequence of this pulsar 
provides further evidence that the periodic nulling originate due to different 
physical mechanism compared to subpulse drifting.

\section{Summary and Conclusion}\label{sec:sum}
\noindent
We have carried out a detailed analysis of the single pulse emission from the 
core-cone Triple pulsar J2002+4050, observed around 1.6 GHz frequencies using 
the Effelsberg radio telescope. The single pulse sequence showed the presence 
of periodic modulations at two different timescales, 2.5$P$ corresponding to 
the phenomenon of subpulse drifting and 40$P$ arising due to periodic nulling. 
Subpulse drifting was seen only in the two conal components but was absent in 
the central core emission. Unlike the majority of pulsars with central line of 
sight traverse of the emission beam, the drifting in this case showed large 
phase variations across both conal components. The periodic modulation features
corresponding to both these phenomena showed variations with time where they 
had sharp features at certain instances which became diffuse at other times. We
have also detected the presence of nulling for the first time in this pulsar. 
The pulsar was in the null state for 11.5 percent of the observing durations. 
We were not able to statistically identify the null pulses due to the low 
sensitivity of detection of single pulses, however, there were regular 
intervals of nulling lasting between 5-10 periods which could be identified by 
visual inspection. These nulling durations were demonstrated to exhibit 
periodic behaviour which was coincident with the 40$P$ modulations seen in the 
pulse sequence. The periodic nulling and subpulse drifting are expected to 
arise due to different physical mechanisms. Our analysis for the pulsar 
J2002+4050 provides further evidence to demonstrate the distinction between the
two phenomena.

\section*{Acknowledgments}
We thank Dr. Ramesh Karuppusamy for his assistance with technical issues during
the observations. We also thank Dr. Karolina Ro\.{z}ko and S\l{}awomir 
Bia\l{}kowski for their assistance with the observations. R.B. thanks Dr. 
Dipanjan Mitra and Prof. George I. Melikidze for useful discussions. This work 
is based on observations with the 100-m telescope of the MPIfR 
(Max-Planck-Institut f\"{u}r Radioastronomie) at Effelsberg. This work has 
received funding from the European Union's Horizon 2020 research and innovation
program under grant agreement No 730562 [RadioNet].

\section*{Data Availability}
The time series single pulse data for PSR J2002+4050~in ascii format underlying
this article will be shared on reasonable request to the corresponding author.

\end{document}